\documentclass[authoryear,review,12pt,fleqn]{elsarticle}
\usepackage[]{graphicx}
\usepackage[]{color}
\definecolor{fgcolor}{rgb}{0.345, 0.345, 0.345}

\usepackage{framed}
\makeatletter
\newenvironment{kframe}{%
 \def\at@end@of@kframe{}%
 \ifinner\ifhmode%
  \def\at@end@of@kframe{\end{minipage}}%
  \begin{minipage}{\columnwidth}%
 \fi\fi%
 \def\FrameCommand##1{\hskip\@totalleftmargin \hskip-\fboxsep
 \colorbox{shadecolor}{##1}\hskip-\fboxsep
     \hskip-\linewidth \hskip-\@totalleftmargin \hskip\columnwidth}%
 \MakeFramed {\advance\hsize-\width
   \@totalleftmargin\z@ \linewidth\hsize
   \@setminipage}}%
 {\par\unskip\endMakeFramed%
 \at@end@of@kframe}
\makeatother

\definecolor{shadecolor}{rgb}{.97, .97, .97}
\definecolor{messagecolor}{rgb}{0, 0, 0}
\definecolor{warningcolor}{rgb}{1, 0, 1}
\definecolor{errorcolor}{rgb}{1, 0, 0}
\newenvironment{knitrout}{}{} % an empty environment to be redefined in TeX

\usepackage{amssymb,amsfonts,amsthm,amsmath,graphicx,url}
\usepackage{lineno}
\usepackage{xr}
\newcommand{\beq}{\begin{eqnarray*}}
\newcommand{\eeq}{\end{eqnarray*}}
\newcommand{\beqq}{\begin{equation}}
\newcommand{\eeqq}{\end{equation}}

\def\balpha{{\bvec \alpha}}
\def\bbeta{{\bvec \beta}}
\def\bgamma{{\bvec \gamma}}

\def\boeta{{\bvec \eta}}

\def\bvec#1{\mbox{\boldmath $#1$}}

\def\bu{{\bvec u}}
\def\bx{{\bvec x}}
\def\bp{{\bvec p}}

\def\bw{{\bvec w}}
\def\bx{{\bvec x}}
\def\bZ{{\bvec Z}}

\usepackage{trackchanges}

\newcommand{\whpsi}{\widehat{\psi}}
\newcommand{\whp}{\widehat{p}}

\newcommand{\whalpha}{\widehat{\balpha}}
\newcommand{\whbeta}{\widehat{\bbeta}}

\newcommand{\whbetas}{\widehat{\beta}\,}

\newcommand*\patchAmsMathEnvironmentForLineno[1]{
  \expandafter\let\csname old#1\expandafter\endcsname\csname #1\endcsname
  \expandafter\let\csname oldend#1\expandafter\endcsname\csname end#1\endcsname
  \renewenvironment{#1}
     {\linenomath\csname old#1\endcsname}
     {\csname oldend#1\endcsname\endlinenomath}}
\newcommand*\patchBothAmsMathEnvironmentsForLineno[1]{
  \patchAmsMathEnvironmentForLineno{#1}
  \patchAmsMathEnvironmentForLineno{#1*}}
\AtBeginDocument{
\patchBothAmsMathEnvironmentsForLineno{equation}
\patchBothAmsMathEnvironmentsForLineno{align}
\patchBothAmsMathEnvironmentsForLineno{flalign}
\patchBothAmsMathEnvironmentsForLineno{alignat}
\patchBothAmsMathEnvironmentsForLineno{gather}
\patchBothAmsMathEnvironmentsForLineno{multline}
}

\addeditor{RH}
\addeditor{NK}

\IfFileExists{upquote.sty}{\usepackage{upquote}}{}
\begin{document}

    \begin{frontmatter}

\title{{\bf Two-stage approaches to the analysis of occupancy data II. The  heterogeneous model and conditional likelihood} \\
%(Analysis of occupancy data)
}

    \author[unimelb]{N.\ Karavarsamis\corref{cor1}}\ead{nkarav$@$unimelb.edu.au}

    \author[unimelb]{ R.\ M.\ Huggins}

    \address[unimelb]{ School of Mathematics and Statistics
   The University of Melbourne, Victoria 3010, Australia. }

   \cortext[cor1]{Corresponding author}

\begin{abstract}
    Occupancy models involve both the probability a site is occupied and the probability occupancy is detected.
The  homogeneous occupancy model, where the occupancy and detection probabilities are the same at each site, admits an orthogonal parameter transformation that yields a two-stage process to calculate the maximum likelihood estimates so that it  is not necessary to simultaneously estimate the
occupancy and detection probabilities.
The two-stage approach is examined here for the heterogeneous occupancy model where the occupancy and detection probabilities now depend on covariates that may
vary between sites and over time. 
There is no longer an orthogonal transformation but this approach effectively reduces the parameter space and allows fuller use of the R functionality. This permits use of existing vector generalised linear models methods to fit models for detection and
allows the development of an iterative weighted least squares  approach to fit models for occupancy. Efficiency is examined in a simulation study and the full maximum likelihood and two-stage approaches are compared on several data sets.\footnote{Software appears as annexes in the electronic version of this manuscript.}
\end{abstract}


     \end{frontmatter}

\section{Introduction}

Occupancy models were introduced in \citet{mac02}.
They model the probability a site is occupied and the probability that occupancy is detected.
We assume here that occupancy of a site is permanent over the observation period.
Data are collected over repeated visits to a number of sites and consist of observations on whether occupancy is detected.
It is common that both the occupancy and detection probabilities are modelled in terms of covariates, which can be time varying.
Occupancy is related to time independent site covariates and detection can be related to both these site covariates and the time varying covariates.
Thus modelling detection can be more complex than modelling occupancy.

Occupancy models are currently fitted to data using the full likelihood where the parameters associated with occupancy and detection are simultaneously estimated. The likelihood may be maximised numerically  using the \citet{r} package \texttt{unmarked} \citep{fiske14} for example.
We have observed that the full likelihood can be numerically unstable. This is distinct from boundary solutions that occur in occupancy models, as noted in
\citet{wintle04,  guillera10,kara13, hutchinson15}.
Without constraints, the full likelihood may not converge, may give
local maxima, or give estimates beyond the boundaries of the parameter space. Using the common logit transformation can still give estimated probabilities that are effectively zero or one.  
We examine this in simulations in Section \ref{sec-sims} where it is seen that maximum likelihood can give extreme estimates of  occupancy parameters when the two-stage approach does not.

Bayesian methods have been developed to estimate occupancy and detectability, for example \citet{milne89,winbugs00,wintle03,mac06, gimenez07,royle08,gimenez09,
mac09,unmarked, martin11, aing11,hui11}. An empirical Bayes method is known but this can underestimate the variance of the posterior distribution \citep{royle08,unmarked}.
Penalized likelihood methods for occupancy have also been developed to help overcome the numerical instability of the maximum likelihood estimators \citep{Moreno:2010ec,hutchinson15}.
These may be fitted using the {\tt occuPEN} and {\tt occuPEN\_CV} functions in {\tt unmarked} package.
In our two-stage approach we address potential instability by considering detection and occupancy
separately. This allows us to compute the estimates over two lower dimension parameter spaces. Moreover, the more complex modelling of the effect of time dependent covariates on the detection
probabilities is relatively straightforward in the two-stage approach.

To help stabilise the numerical optimization algorithm the package \texttt{unmarked} (see p.\ 2 of the R vignette
\citet{fiske14}\footnote{\url{cran.at.r-project.org/web/packages/unmarked/vignettes/unmarked.pdf}}) recommends that covariates be standardized.  However, as observed in the documentation for the \texttt{unmarked} package, standardizing may cause
problems with standard R functions, such as \texttt{predict}. The {\tt smartpred} package may solve some issues but not in all instances of data dependent parameters. Moreover, there is no guarantee that users will standardise their data. In addition, the choice of algorithm for the numerical maximisation may be changed in {\tt optim}, however this may be sensitive to the algorithm used.

Recently \citet{kara17} showed that for the homogeneous occupancy model a simple transformation yielded orthogonal parameters resulting in a two-stage estimation procedure that simplified the computation of
the estimates. We see in Section \ref{sec-two-stage} that this no longer holds for heterogeneous models.
Following the homogeneous case,  a conditional likelihood is used to estimate detection probabilities which is the first stage of the analysis.
This may be implemented using the {\tt vglm} function in the {\tt VGAM} package in R \citep{yee10,yee2015,Yee:2014vg}.
 In the second stage, the remaining partial likelihood, evaluated at the estimated detection
probabilities from the first stage, is used to estimate the occupancy probabilities. This effectively reduces the parameter space  and allows the use of vector generalized linear model methods to fit models for detection. The partial likelihood for occupancy may be maximised using several numerical methods, here we implement this using an iterative weighted least squares (IWLS) approach.

Our notation is given and the full likelihood is examined in Section \ref{sec-notat}. In Section \ref{sec-two-stage} we describe the two-stage approach.
In the first stage in Section \ref{sec-detect} we use conditional likelihood to estimate the detection probabilities,
with time independent detection probabilities discussed in Section \ref{sec-tid}
and time dependent detection probabilities considered in Section \ref{sec-td0}.
In the second stage in Section \ref{sec-est_occup} we introduce a partial likelihood approach to estimate the occupancy probabilities using the detection probabilities estimated in the first stage. In Section \ref{sec-iter1} we give an IWLS algorithm to compute the
occupancy estimates.  A  simulation study is conducted in Section \ref{sec-sims} and the methods are applied to several data sets in Section \ref{sec-applic}.
Some discussion is given in Section \ref{sec-disc}. Some technical derivations and the implementation of {\tt vglm} in this setting are given in the appendices.

\section{Notation and Full Likelihood}\label{sec-notat}

Consider $S$ sites labelled $s=1,\dots,S$ and $\tau$ occasions at each site where the presence of a species may be observed.
We suppose that occupancy is constant over the observation period.
Let $\psi_s$ be the probability that site $s$ is occupied and $p_{sj}$ be the
probability the species is observed at site $s$ on  visit $j$ given it is present at site $s$. Then $\theta_s=1-\prod_{j=1}^\tau(1-p_{sj})$ is the probability of at least one detection at site $s$ given the site is occupied. If there is no dependence on the visit then $p_{sj}=p_s$
and $\theta_s=1-(1-p_s)^\tau$.
Let $Y_{sj}$ take the value 1 if an individual was detected at site
$s$ on occasion $j$ and zero otherwise.
Let $Y_s=\sum_{j=1}^\tau y_{sj}$ denote
the number of occasions upon which the species was detected at site $s$. We let $Z_s=I(y_s=0)$ be the indicator of no detections at site $s$. Reorder the $S$ sites $s=1,\dots,O, O +1, \ldots S$, where $1, \ldots, O$ denote the sites at which at least one detection occurred and $O +1, \ldots S$ the remainder sites at which no sightings
occurred.

It is common for covariates that may be related to detection or occupancy to be associated with each site.
Suppose that $\psi_s = h(x_s^T \balpha)$ where  $x_s$ is a vector of covariates associated with site $s$ and $\balpha \in \mathbb{R}^p$ is a vector of coefficients.
 Let $p_{sj}$ be the
probability of detection at site $s$ on occasion $j$ if site $s$ is occupied.
We take $p_{sj}=h(u_{sj}^T\bbeta)$
where $u_{sj}$ is  a vector of covariates associated with site $s$ on occasion $j$,
and let $\theta_s=1-\prod_{j=1}^\tau(1-p_{sj})$, for site and time dependent detection.
For time independent detection probabilities, we write $p_{sj}=p_s= p(u_s, \bbeta) = h(u_s^T\bbeta)$, $j=1,\dots,\tau$,  for a possibly
different vector of covariates $u_s$  to that above and with corresponding coefficient vector $\bbeta \in \mathbb{R}^q$. In most applications $h$ will be the logistic function
$h(x)=(1+\exp(-x))^{-1}$. Let $\bp_s=(p_{s1},\dots, p_{s\tau})^T$ in the time dependent case and $\bp_s=p_s$ otherwise.

The contribution to the full likelihood of site $s$, can be written as
\begin{equation}
L_s(\psi_s,\bp_s)= (1-\psi_s\theta_s)^{z_s} \left\{\psi_s\prod_{j=1}^\tau  p_{sj}^{y_{sj}}(1-p_{sj})^{1-y_{sj}}\right\}^{1-z_s}. \label{lik}
\end{equation}
The full likelihood is based on maximising the product of (\ref{lik}) over all the sites. Let
\begin{align*}
   \ell(\psi_s,\bp_s)&= z_s \log(1-\psi_s\theta_s) + (1-z_s)\log(\psi_s)\\
        &+ (1-z_s)\sum_{j=1}^\tau y_{sj}\log(p_{sj})+(1-z_s)\sum_{j=1}^\tau (1-y_{sy})\log(1-p_{sj})
\end{align*}
be the contribution of site $s$ to the log-likelihood.
Then, assuming sites are independent, the full log-likelihood is $\ell(\balpha,\bbeta)=\sum_{s=1}^S\ell(\psi_s,\bp_s)$.

\section{The Two-Stage Approach}\label{sec-two-stage}

Following the homogeneous case of \citet{kara17} an alternate approach is to let $\eta_s$ ($=\psi_s\theta_s$) be the unconditional probability the species is detected at site $s$.
Then we may write the contribution of site $s$ to the full likelihood, (\ref{lik}),  as
\begin{align}
    L(\eta_s,\bp_s)& = (1-\eta_s)^{z_s} \eta_s^{1-z_s} \times\left\{\frac{\prod_{j=1}^\tau p_{sj}^{y_{sj}}(1-p_{sj})^{1-y_{sj}}}{\theta_s}\right\}^{1-z_s} \nonumber \\ %\label{lik-0}  \\
    &=L_1(\eta_s)L_2(\bp_s).\label{lik-1}
    \end{align}
    The partial likelihood component $L_1(\eta_s)$ is a Bernoulli likelihood corresponding to the detection of the species at site $s$ and the partial component $L_2(\bp_s)$
    is the conditional likelihood
        of $\bp_s$ given at least one detection at the site.
    The contribution of site $s$  to the log-likelihood is then
    \begin{align}
       \ell(\eta_s,\bp_s)&= z_s \log(1-\eta_s) + (1-z_s)\log(\eta_s)\label{eq-eta}  \\
            &+ (1-z_s)\left\{\sum_{j=1}^\tau y_{sj}\log(p_{sj})+\sum_{j=1}^\tau (1-y_{sj})\log(1-p_{sj})
             -\log(\theta_s)\right\}.\label{eq-p}
    \end{align}
    Our interest is in exploiting the decomposition (\ref{lik-1}) to simplify the calculations for complex models. To achieve this, we use (\ref{eq-p})
    to estimate $\bbeta$ in the first stage of the estimation process. Let $\widehat\bbeta$ be the resulting conditional likelihood estimator of $\bbeta$,
      and denote its large sample variance by $V_\beta$.
    In the second stage, let $\widehat\bp_s$ be the fitted value of $\bp_s$
    and $\widehat\theta_s$ the fitted value of $\theta_s$. We then replace $\eta_s$ by $\widetilde\eta_s=\psi_s\widehat\theta_s$ in the
	 log-partial likelihood (\ref{eq-eta}) and maximise this to estimate $\balpha$.

    From (\ref{eq-p})
    the conditional likelihood estimator of $\bbeta$ arises from solving
    \begin{equation}
    0=S_1(\bbeta)=\sum_{s=1}^S \frac{\partial \ell(\eta_s,\bp_s)}{\partial \bp_s^T} \frac{\partial \bp_s^T}{\partial\bbeta}.\label{eq-beta}
    \end{equation}
    Rather than solving (\ref{eq-beta}) the maximum likelihood estimators of $\bbeta$ arise from solving
    \begin{align}
    0&= \sum_{s=1}^S  \frac{\partial \ell(\eta_s,\bp_s)}{\partial\bbeta}\nonumber \\
    &= \sum_{s=1}^S \left\{ \frac{\partial \ell(\eta_s,p_s)}{\partial\eta_s}  \frac{\partial \eta_s}{\partial \bp_s^T}
    + \frac{\partial \ell(\eta_s,p_s)}{\partial \bp_s^T} \right\} \frac{\partial \bp_s^T}{\partial \bbeta} \nonumber \\ % \label{eq-lik-p}\\
    &= \sum_{s=1}^S  \frac{\partial \ell(\eta_s,p_s)}{\partial\eta_s} \frac{\partial \eta_s}{\partial \bp_s^T}+S_1(\bbeta). \nonumber
    \end{align}
    Thus unlike the simple homogeneous model considered in  \citet{kara17} the conditional likelihood estimators will not be the mle's.

\subsection{Stage 1: Estimating the Detection Probabilities}\label{sec-detect}

 For the homogeneous model \citet{kara17} included in their supplementary materials a plot of the estimated occupancy probability against values of the detection probability.
 This plot suggests that the occupancy probability is relatively insensitive to small changes in the detection probability. Thus modelling the detection probability may not be crucial.
 However, particularly if there are big changes in the detection probabilities between visits, correct modelling will be expected to improve the estimates and may be of interest independent to $\psi$.

  \subsubsection{Time Independent Detection Probabilities}\label{sec-tid}

  This is the simplest case, apart from constant detection probability.
In this case $\bp_s=p_s$ and the conditional likelihood
reduces to
\[
L_2(\bbeta)=\prod_{s=1}^O\frac{ p_s^{y_s}(1-p_s)^{\tau-y_s}}{\theta_s},
\]
which is a function of  the number of detections at each site where there was at least one detection, i.e. $y_s$, $s=1,\dots,O$.
  This is the conditional likelihood of \citet{Huggins:1989hb} which may be easily maximised using the {\tt VGAM} package, with nomenclature similar to that used in generalised linear models  \citep[\S 17.2 ][]{yee2015,Yee:2014vg}.
 See \ref{vgam}, \ref{vgam-tid} and Section \ref{sec-applic} for details and examples.

\subsubsection{Time Dependent Detection Probabilities}\label{sec-td0}

Recall that in this case we have distinct probabilities $p_{sj}$, $j=1, \ldots, \tau$ for different visits to site $s$.
  A simple extension of the time independent model allows an effect  of the $j$th visit in the
model for $p_{sj}$. That is, the covariate vector $u_{sj}$ contains an indicator of the visit time.
This is modelled by allowing the intercept to vary with the visit,
and this is easily implemented in the {\tt VGAM} package.  See \ref{vgam-t} and  Section \ref{sec-applic}.
More generally environmental variables such as temperature or the time of day the visit was conducted
may vary between visits.
When we allow the detection probabilities to depend on time dependent covariates, the
conditional likelihood corresponding to a site $s$ where occupancy was detected is now
\begin{equation*}
L_2(\bp_s)=\frac{\prod_{j=1}^\tau p_{sj}^{y_{sj}}(1-p_{sj})^{1-y_{sj}}}{\theta_s}. %\label{eq-cond-probs}
\end{equation*}
That is, the detections
form a sequence of independent Bernoulli trials but we only observe the outcome if there is at least one detection. Again this model can be fitted using the {\tt VGAM} package. See \ref{vgam-t} for details.

\subsection{Stage 2: Estimating Occupancy Probabilities}\label{sec-est_occup}

To estimate $\balpha$ we maximise the partial likelihood
$
\prod_{s=1}^S L_{1s}(\widetilde\eta_s)
$ where, as noted above,  $\bp_s$ and hence $\theta_s$ has been replaced by its estimator from the first stage $\whp_s=\bp_s(\widehat \bbeta)$.
The partial likelihood is
$
L_1(\balpha)=\prod_{s=1}^S L_{1s}(\widetilde\eta_s)
\propto\prod_{s=1}^S
(1-\psi_s\widehat\theta_s)^{z_s} \psi_s^{1-z_s}$. Let $w_s = 1 - z_s$, then the log-partial likelihood is
\begin{equation}
    \ell(\balpha)=\sum_{s=1}^S \left\{(1-w_s)\log(1-\psi_s\widehat\theta_s)+w_s\log(\psi_s)\right\}.\label{eq-log-PL}
\end{equation}
This may be maximised numerically using the {\tt optim} function in R (referred to as ``Partial'' in tables). However, there are two other possible approaches.

 \subsubsection{Iterative Weighted Least Squares}\label{sec-iter1}

  An alternative to this method that is commonly used to compute estimates from generalised linear models is the well known iterative weighted least squares (IWLS)
 approach. To define this estimator for a logistic model, let  the matrix $X$ have $s$th column $x_s$. Let  $\bw=(w_1,\dots,w_S)^T$, $\eta_s=\theta_s\psi_s$, $\boeta=(\eta_1,\dots,\eta_S)^T$. Let  $\boeta(\balpha)$ be $\boeta$
 evaluated at $\balpha$.
 Set $V={\rm diag}\{(1-\boeta)\boeta\}$ and   $U={\rm diag}\{\theta_s\psi_s(1-\psi_s)\}$. Let $\balpha^{(k)}$ be the estimate at the $k$th step and
 let $\bZ=UX\balpha^{(k)}+ \bw-\boeta(\balpha^{(k)})$. Then the estimate at the $(k+1)$th is
 $
 \balpha^{(k+1)}=\left(X U V^{-1}U X^T \right)^{-1}X U V^{-1} U\bZ$.
The IWLS estimate is obtained by repeating this step until convergence. Details are given in \ref{sec-iwls}. An estimate of the expected Fisher information corresponding to the partial likelihood, $E\left\{I(\balpha,\bbeta)\right\}$, is given by $\widetilde I(\balpha,\bbeta)=X U V^{-1}U X^T$.

\subsubsection{Iterative Offset}

As $\theta_s$ does not depend on $\balpha$, maximising (\ref{eq-log-PL})
is equivalent to maximising
\begin{equation*}
    \ell(\balpha)=\sum_{s=1}^S \left\{(1-w_s)\log(1-\widehat\eta_s)+w_s\log(\widehat\eta_s)\right\}. %\label{eq-log-PL2}
\end{equation*}
where $\widehat\eta_s-\psi_s\widehat\theta_s$.
  Let
$
a_s (x_s)=\log(\widehat\theta_s)-\log\{1+\exp(\alpha^T x_s)(1-\widehat\theta_s)\}
$. Then under the logistic model
\begin{align*}
\eta_s=\psi_s\widehat\theta_s&={\exp(\alpha^T x_s+\log(\widehat\theta_s))}/{\{1+\exp(\alpha^T x_s)\}}\\
&= {\exp(\alpha^T x_s+a_s(x_s))}/{\{1+\exp(\alpha^T x_s+a_s(x))\}}
\end{align*}
 and $a_s(x_s)$ has the appearance of an offset. However, it is a function
of the linear predictor $\alpha^T x_s$. This allows an alternative iterative
approach.

\subsection{Estimating the Standard Errors}

We give the asymptotic variances
of our occupancy estimators in the linear logistic case.
Denote the partial score function by $Q(\alpha,\beta)=\partial
\log\left(L_1(\balpha,\bbeta)\right)/\partial \balpha$,
$I(\balpha,\bbeta)=-\partial Q(\balpha,\bbeta)/\partial \balpha^T$ and
$\widetilde B(\balpha,\bbeta)=-\partial Q(\balpha,\bbeta)/\partial \bbeta^T$.
Let
$\whalpha(\bbeta)$ be the estimator of $\balpha$  arising from solving $Q(\balpha,\bbeta)=0$
for a given
$\bbeta$. We show in \ref{sec-glmp} that under mild regularity conditions an estimator of the variance of $\whalpha(\whbeta)$ is
\begin{align}
\hspace*{-10mm} \widehat{\rm Var}\{\whalpha(\whbeta)\}  & = \label{eq-se}
 \\
& \hspace*{-15mm}  I\{\whalpha(\whbeta),\whbeta\}^{-1}
 +I\{\whalpha(\whbeta),\whbeta\}^{-1}
\widetilde B\{(\whalpha(\whbeta),\whbeta)\}
\widehat V_\beta \widetilde B\{\whalpha(\whbeta),\whbeta\}^T
I\{\whalpha(\whbeta),\whbeta\}^{-1}. \nonumber
\end{align}
The estimated occupancy probability is $\whpsi_s=[1+\exp\{-x_s^T \whalpha(\whbeta)\}]^{-1}$, which
is easily seen to have the  estimated approximate variance
$
\widehat{\rm Var}(\whpsi_s)=\{\whpsi_s(1-\whpsi_s)\}^2
x_s^T \widehat{\rm Var}\{\whalpha(\whbeta)\} x_s$.

To compute (\ref{eq-se}) note that in both the time homogeneous and inhomogeneous cases we have
\begin{equation}
I(\balpha,\bbeta)=
\sum_{s=1}^S x_sx_s^T\left\{
\frac{\theta_s-2\psi_s(\balpha)\theta_s +
\psi_s(\balpha)^2\theta_s^2+w_s(1-\theta_s)
}{\{(1-\psi_s(\balpha)\theta_s\}^2}\right\}
\psi_s(\balpha)\{1-\psi_s(\balpha)\}.\label{eq-I}
\end{equation}
However, $\widetilde B(\balpha,\bbeta)$ is computed differently.
In the time homogeneous  case we have
\begin{equation}
\widetilde B(\balpha,\bbeta)=-\frac{\sum_{s=1}^S x_su_s^T\psi_s(1-\psi_s)(1-w_s) \tau(1-\theta_s)p_s}
{(1-\psi_s\theta_s)^{2}},\label{eq-B-homog}
\end{equation}
whereas in the time heterogeneous case we have
\begin{equation}
\widetilde B(\balpha,\bbeta)=-\sum_{s=1}^S
  \frac{x_s\psi_s(1-\psi_s)(1-w_s)
    (1-\theta_s) \sum_{j=1}^\tau p_{sj} u_{sj}^T}{(1-\psi_s\theta_s)^2}.\label{eq-B-hetero}
\end{equation}
Then in either case, the expression \eqref{eq-se} holds (see \ref{sec-glmp}).
Note that if $p_{sj}\equiv p_s$ and $u_{sj}\equiv u_s$ then
$ \sum_{j=1}^\tau p_{sj} u_{sj}^T=\tau p_s u_s$  (\ref{eq-B-homog}) and (\ref{eq-B-hetero}) are the same. In computing  $\widetilde B(\balpha,\bbeta)$ we can replace $w_s$ by
its expectation $\psi_s\theta_s$.

\section{Simulations}\label{sec-sims}

To evaluate the two-stage approach (described in Section~\ref{sec-est_occup}) we used one site covariate for occupancy, an independent site covariate for detection and a further independent single time varying covariate
that also varied between sites. These were all taken to have standard normal distributions, reflecting that it is common to standardise the covariates.
We took varying parameter values to reflect different mean occupancy and detection probabilities. For each value of $S$ and $\tau$ we simulated the covariates once then for varying parameter values simulated
occupancy and detection. We first conducted simulations to determine which of the three methods 
%, Partial (direct maximisation using {\tt optim}), IWLS or Iterative offset, 
of estimating occupancy performed best.
Firstly we conducted simulations to determine which of the methods, IWLS, direct maximisation of the partial likelihood with {\tt optim} (Partial) or iterative offset (Iterative) to use.
Direct maximisation using {\tt optim} allows different optimization methods, the option BFGS numerical maximisation method is adopted here. Comparison to CG and Nelder-Mead method displayed  improved convergence for BFGS.
We consider $S=500$ and $\tau=5$, $\balpha=(1,1)$, and $\bbeta=(-1.5,-0.5,-0.5)$ giving for our simulated covariates an average occupancy probability of 0.70 and an average probability of detection at least once at a site of 0.65. We took 1000 simulations. The results are in Table \ref{tab-m1}. Efficiencies are computed relative to the partial likelihood and are computed using the usual ratio of the variances (Efficiency) and the ratio of median absolute deviations squared (Efficiency (mad)). Median absolute deviations (mad) are given by mad $= c \times$ median $(|x_i - x_M|)$, where $c = 1/\Phi(-1)(3/4)$, $\Phi(-1)(3/4)$ is the third quartile of the inverse standard normal distribution, and $x_M$ is the median of $x_i$. We used the function {\tt mad} in R to calculate these. The medians showed little bias in any of the methods although when the means were computed, direct maximisation of the partial likelihood (Partial) exhibited considerable bias. The median absolute deviations of the IWLS and Partial methods were similar as is also evident in the efficiencies computed using the mad.
With a smaller number of site visits, $\tau=3$ in Table \ref{tab-m2} there is now some evidence of bias in all methods and the IWLS is clearly the most efficient. In this case the average probability of detection at least once at a site was 0.47.

\begin{table}[ht]
\centering
    \caption{Stage 2: Efficiency for estimating occupancy probabilities  for the two-stage approach for three methods; IWLS, direct maximisation of the partial likelihood using {\tt optim} (Partial), or iterative offset (Iterative). Simulation results to compare numerical methods $S=500$, $\tau=5$, mean $\psi=0.7$, mean $\theta= 0.65$, $\balpha=(1,1)$ and $\bbeta= (-1.5, -0.5, -0.5)$.}
    \begin{small}
\begin{tabular}{l rr rr rr}
  %\hline
&\multicolumn{2}{c}{IWLS}&\multicolumn{2}{c}{Partial}&\multicolumn{2}{c}{Iterative}\\
  \hline
Actual $\alpha$ & 1.00 & 1.00 &  1.00 &  1.00 &  1.00 &  1.00 \\ %\hline
  Median & 1.02 & 1.00 & 1.02 & 1.00 & 1.02 & 1.01 \\
  mad & 0.25 & 0.25 & 0.25 & 0.26 & 0.32 & 0.31 \\ %\hline
  Mean & 1.03 & 1.02 & 1.97 & 1.47 & 1.15 & 1.15 \\
  sd & 0.28 & 0.26 & 5.66 & 2.70 & 1.03 & 1.04 \\ %\hline
  Efficiency & 41656.85 & 10910.59 &  &  & 3014.89 & 677.40 \\
  Efficiency(mad) & 100.00 & 107.35 &  &  & 61.96 & 67.59 \\
   \hline
\end{tabular}
\end{small}
\label{tab-m1}
\end{table}

\begin{table}[ht]
\centering
    \caption{Stage 2: Efficiency for estimating occupancy probabilities for the two-stage approach for three methods; IWLS, direct maximsiation of the partial likelihood using {\tt optim} (Partial), or iterative offset (Iterative). Simulation results to compare numerical methods $S=500$, $\tau=3$, mean $\psi=0.7$, mean $\theta= 0.47$, $\balpha=(1,1)$ and $\bbeta= (-1.5, -0.5, -0.5)$.}
    \begin{small}
 \begin{tabular}{lrrrrrr}
   %\hline
 &\multicolumn{2}{c}{IWLS}&\multicolumn{2}{c}{Partial}&\multicolumn{2}{c}{Iterative}\\
   \hline
Actual $\alpha$ & 1.00 & 1.00 & 1.00 & 1.00  & 1.00 & 1.00  \\ %\hline
  Median & 1.17 & 0.99 & 1.17 & 1.02 & 1.23 & 1.03 \\
  mad & 0.57 & 0.24 & 0.62 & 0.31 & 0.68 & 0.54 \\ %\hline
  Mean & 1.17 & 1.05 & 1.34 & 1.15 & 1.81 & 1.67 \\
  sd & 0.49 & 0.33 & 0.99 & 0.55 & 2.03 & 1.77 \\ %\hline
 Efficiency  & 400.97 & 284.07 &  &  & 23.69 & 9.65 \\
 Efficiency(mad) & 120.18 & 165.64 &  &  & 81.85 & 32.49 \\
   \hline
\end{tabular}
\end{small}
\label{tab-m2}
\end{table}

Next, we compare the IWLS method for the two-stage approach of estimation of occupancy to the full likelihood of \citet{mac02}. In Table \ref{tab-5005} we consider $S=500$ and $\tau=5$. We took 1000 simulations at each parameter combination. With a small number of standardised covariates the MacKenzie maximum likelihood  estimators were expected to perform well. These were computed using the {\tt occu} function in the R package {\tt unmarked} (see \ref{sec-occu} for a brief description).
We report the median and mad of the estimates. With lower detection probabilities occasionally the IWLS algorithm did not converge in 200 iterations. In that case the partial likelihood could be directly maximised.
The bias of both procedures was low. As expected the efficiencies of estimating the parameters associated with detection were low. The efficiencies of the two-stage estimator %\add[NK]{(using IWLS or direct maximisation if IWLS did not converge in 200 iterations)} 
in estimating the occupancy
  probabilities was good and was generally around 100\% for the covariate term, but less that 100\% for the intercept. The large efficiencies for smaller occupancy and detection probabilities were due to
  several unusually large maximum likelihood estimates. Similarly the small efficiencies for smaller detection probabilities but larger occupancy probabilities were due to large values of the two-stage estimates.
  
  \begin{table}[ht]
\centering
    \caption{Simulation results for $S=500$, $\tau=5$ for four studies with mean probabilities $(\psi, p)$: $(0.7, 0.65)$, $(0.31, 0.65)$, $(0.7, 0.37)$ and $(0.31, 0.37)$. Efficiency is based on the variance rather than the mad. Medians for the two-stage method (med two-stage)  and full maximum likelihood estimates (med mle), as well as their median absolute deviations `mad two-stage' and `mad mle', respectively. Occupancy for the two-stage approach estimated with  IWLS method.} %, or direct maximisation if IWLS did not converge.}}
    \begin{tiny}
\begin{tabular}{lrrrrrrrrrr}
%    \hline
  %  &\multicolumn{10}{l}{$S=500$, $\tau=5$}\\
    %\hline
    &\multicolumn{2}{c}{Occupancy}&\multicolumn{3}{c}{Detection}&\multicolumn{2}{c}{Occupancy}&\multicolumn{3}{c}{Detection}\\
  %\hline
  & Int & $x_1$ &Int & $x_2$ & time & Int & $x_1$ &Int & $x_2$ & time \\ \hline
Mean Prob & 0.70 &  & 0.65 &  &  & 0.31 &  & 0.65 &  &  \\ %\hline

    Actual & 1.00 & 1.00 & -1.50 & -0.50 & -0.50 & -1.00 & 1.00 & -1.50 & -0.50 & -0.50 \\ %\hline
    med two-stage & 1.01 & 1.01 & -1.51 & -0.51 & -0.50 & -0.99 & 1.01 & -1.52 & -0.50 & -0.50 \\
    mad two-stage & 0.26 & 0.23 & 0.11 & 0.10 & 0.07 & 0.19 & 0.18 & 0.16 & 0.15 & 0.11 \\ %\hline
    med mle & 1.01 & 1.01 & -1.51 & -0.50 & -0.49 & -0.99 & 1.01 & -1.52 & -0.51 & -0.50 \\
    mad mle & 0.25 & 0.23 & 0.09 & 0.07 & 0.07 & 0.18 & 0.18 & 0.14 & 0.11 & 0.10 \\ %\hline
    Efficiency  & 90.65 & 101.74 & 66.04 & 52.03 & 93.13 & 93.67 & 99.41 & 69.72 & 54.69 & 91.09 \\
    Efficiency(mad) & 91.64 & 98.57 & 70.62 & 54.81 & 94.00 & 93.19 & 102.40 & 73.01 & 49.57 & 96.52 \\
   \hline %\hline
 Mean Prob  & 0.70 &  & 0.37 &  &  & 0.31 &  & 0.37 &  &  \\ %\hline
% & Int & $x_1$ &Int & $x_2$ & time & Int & $x_1$ &Int & $x_2$ & time \\ \hline
   Actual & 1.00 & 1.00 & -2.50 & -0.50 & -0.50 & -1.00 & 1.00 & -2.50 & -0.50 & -0.50 \\ %\hline
   med  two-stage & 1.01 & 0.99 & -2.53 & -0.50 & -0.50 & -0.97 & 1.03 & -2.55 & -0.52 & -0.50 \\
   mad  two-stage & 0.65 & 0.41 & 0.23 & 0.19 & 0.10 & 0.41 & 0.27 & 0.37 & 0.29 & 0.15 \\ %\hline
   med mle & 1.03 & 1.00 & -2.51 & -0.50 & -0.50 & -0.97 & 1.05 & -2.53 & -0.49 & -0.50 \\
   mad mle & 0.57 & 0.39 & 0.16 & 0.09 & 0.09 & 0.36 & 0.28 & 0.28 & 0.16 & 0.14 \\ %\hline
  Efficiency & 0.15 & 0.22 & 43.13 & 23.13 & 81.39 & 371.60 & 3223.71 & 50.28 & 23.03 & 81.91 \\
   Efficiency(mad) & 78.85 & 87.63 & 51.67 & 22.50 & 84.41 & 78.39 & 102.95 & 57.02 & 29.27 & 85.60 \\
  \hline
\end{tabular}
\end{tiny}
\label{tab-5005}
\end{table}

\begin{figure}[ht!]
\centering
\includegraphics[scale=0.6]{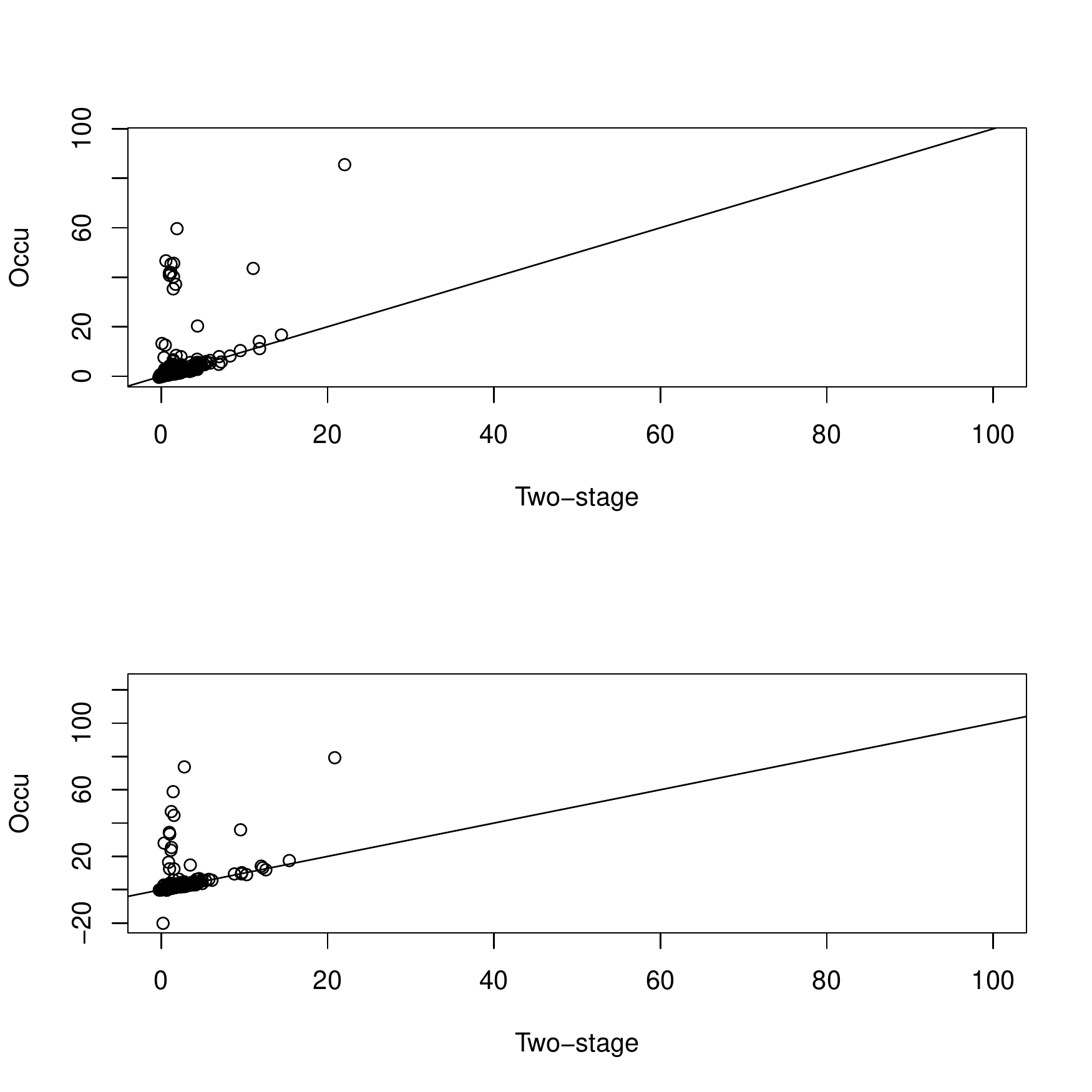}
\caption{Comparison of estimated occupancy parameters ($\hat\balpha$) between {\tt occu} and two-stage method with IWLS for 1000 simulations with $\balpha=(1,1)$, $\bbeta=(-1.5, -0.5)$, and $\beta_t= -0.5$. Top figure shows intercept estimates and bottom figure estimates for the slope parameter.}
\label{eff-plot}
\end{figure}

A plot of 1000 simulated occupancy estimates from our approach were compared to the full likelihood (see Figure \ref{eff-plot}).
The model included time varying covariates using the two-stage approach for the IWLS method and the same was modelled with {\tt occu}; $\balpha=(1,1)$, $\bbeta=(-1.5, -0.5)$, and $\beta_t= -0.5$. Overall, the two-stage estimator gave estimates that were more accurate and more consistent and that {\tt occu} may give extreme estimates of occupancy parameters when the two-stage approach does not. There was a single outlier that was omitted from the plot for clarity. Corresponding summary statistics are given in Table \ref{tab-comps}.  

Table \ref{tab-agr} shows that {\tt occu} gives estimates that are large four times more often than our approach. We present agreement between our approach and {\tt occu} to estimating occupancy. 
Agreement between the two methods is defined as the number of estimates that are either both or neither greater than three ($\hat\alpha_1 > 3$), less than or equal to three ($\hat\alpha_1 \leq 3$), or when these  disagree. {\tt occu} gives estimates that are greater than three (i.e $\hat\alpha_1 >3$) four times more often than our IWLS method i.e. 36 to 12 (Table \ref{tab-agr}). The table clearly demonstrates there is no universal best method for finding estimates for occupancy. When IWLS fails i.e. does not converge, then we recommend using {\tt optim} or {\tt occu}.

\begin{table}[ht]
\centering
\caption{Simulation study of 1000 estimates  of the occupancy parameters with $\balpha=(1,1)$, $\bbeta=(-1.5, -0.5)$, and $\beta_t= -0.5$.}
 \begin{small}
\begin{tabular}{lrrrrr}
  %\hline
  &  \multicolumn{2}{c}{Occupancy} &  \multicolumn{3}{c}{Detection} \\
 & Int & $x_1$ & Int & $x_2$ & time \\ 
  \hline
Mean Prob & 0.72 &  & 0.63 &  &  \\ 
  Actual & 1.00 & 1.00 & -1.50 & -0.50 & -0.50 \\ 
  med two-stage & 0.98 & 1.03 & -1.53 & -0.52 & -0.51 \\ 
  mad two-stage & 0.63 & 0.57 & 0.25 & 0.28 & 0.15 \\ 
  sd two-stage & 7.82 & 9.94 & 0.28 & 0.28 & 0.16 \\ 
  med mle & 1.10 & 1.16 & -1.52 & -0.53 & -0.51 \\ 
  mad mle & 0.70 & 0.65 & 0.21 & 0.17 & 0.15 \\ 
  Efficiency & 67.55 & 45.33 & 68.41 & 38.92 & 90.47 \\ 
  Efficiency(mad) & 99.94 & 125.16 & 71.13 & 38.25 & 93.37 \\ 
   \hline
\end{tabular}
 \end{small}
 \label{tab-comps}
\end{table}

\begin{table}[ht]
\centering
\caption{Agreement for intercept estimates greater, or less, than three when the actual value to be estimated is $\alpha_1 =1$.}
\begin{small}
\begin{tabular}{rrr}
  %\hline
  & \multicolumn{2}{c}{{\tt occu} method} \\
Two-stage IWLS &  $\hat\alpha_1 \leq 3$ & $ \hat\alpha_1 > 3$ \\ 
  \hline
 $\hat\alpha_1 \leq 3$ & 832 &  36 \\ 
   $\hat\alpha_1 > 3$  &  12 &  37 \\ 
   \hline
\end{tabular}
\end{small}
\label{tab-agr}
\end{table}

\section{Applications}\label{sec-applic}

\subsection{Data Set 1}
\citet{hutchinson15} use a publicly available data set from
\citet{dryad_t40f2}\footnote{https://datadryad.org//resource/doi:10.5061/dryad.t40f2} to illustrate their penalized likelihood approach.
 The data set contains detections of 25 avian species over 3 visits to each of 656 sites from a field study in 2011 in southern Indiana, USA. We use the entire data set. There are six site-specific vegetation covariates available
(labelled {\tt vegcov1, vegcov2, \dots vegcov6}) and
four time dependent survey covariates  {\tt time}, {\tt temp}, {\tt cloud} and {\tt julian} measured for each visit to each site.
    These are included in the data data frame as {\tt time1, time2, \dots julian2, julian3}.
All the covariates have been standardised.
 We consider the entire data set for the first species.
 We considered four models for the detection probabilities. The first only involved the site covariates, the second the site covariates and time dependent intercepts,
 the third site and time dependent survey covariates and the last  involved site and time dependent survey covariates and time dependent intercepts. The values of the AIC from the conditional likelihood are:
Site only: AIC = 1517.4, Site + Time Varying Intercept: AIC = 1511.1, Site + Survey: AIC = 1488.3, Site + Survey + Time Varying Intercept: AIC = 1486.2. The best model of these  includes the site covariates, time dependent survey covariates and time dependent intercepts. The resulting two-stage (Two-stage) estimates (with IWLS method) 
%or direct maximisation if IWLS did not converge)} 
are displayed in Table \ref{tab-modelt} along with the full maximum likelihood estimates (Full likelihood) computed using the {\tt occu} function in the R package {\tt unmarked} (fitting the model with {\tt occu} is briefly described in~\ref{sec-occu}). The maximum likelihood and two-stage estimates are very similar. %\footnote{{\color{blue}What method used in Table 3 for psi.hat: IWLS, Partial or iterative, or combination????}}

In the model for detection there are 6 site covariates and 4 survey covariates. This gives $2^{10}=1024$ possible models (or $2^{11}$ if one allows time varying intercepts.) Whilst this is a large number of models, in the absence of variable selection methods in {\tt VGAM} it is nevertheless feasible to compute the AIC for each model. We can then repeat this process for the model for occupancy after fixing the best detection model. The best fitting model for the detection probabilities using the AIC is indicated in Table \ref{tab-modelt}. The function {\tt vglm} allows more flexibility in the modelling. For example,
%by changing {\tt parallel.t=FALSE$\sim$0} to {\tt parallel.t=TRUE$\sim$0}
by changing {\tt parallel.t=FALSE$\sim$0} to {\tt parallel.t=TRUE$\sim$0}
the coefficients associated with each variable in the detection model may be time varying. We do not pursue this further here. Of course as occupancy is assumed constant over the visits we do not model
the occupancy coefficients as time varying.

\begin{table}[!ht]
    \caption{Occupancy and detection estimates for full likelihood and two-stage approaches for the detection model with site and survey covariates and time varying intercept for the Hutchinson data
    ($^*$ indicates the variables retained in the best fitting model using the two-stage approach). For each covariate, we report its: estimate (Estimate), standard error (se), Student's \emph{t}-statistic ($t$), and $p$-value ($p$).}
\centering
\begin{tiny}
\begin{tabular}{lrrrrrrrr}
    \hline
    &\multicolumn{4}{c}{Full Likelihood}&\multicolumn{4}{c}{Two-stage}\\ %\hline
%  \hline
 Parameter&  Estimate & se & $t$&$p$ &  Estimate & se & $t$&$p$ \\
  \hline
   \multicolumn{9}{c}{Occupancy $\psi$ }\\ %\hline
Intercept$^*$ & 2.27 & 0.15 & 14.98 & 0.00 & 2.26 & 0.15 & 14.91 & 0.00 \\
  vegcov1$^*$  & 0.50 & 0.18 & 2.78 & 0.01 & 0.52 & 0.17 & 3.00 & 0.00 \\
  vegcov2 & 0.03 & 0.17 & 0.15 & 0.88 & 0.01 & 0.17 & 0.04 & 0.97 \\
  vegcov3 & 0.07 & 0.17 & 0.41 & 0.68 & 0.06 & 0.16 & 0.34 & 0.73 \\
  vegcov4$^*$  & 0.37 & 0.13 & 2.89 & 0.00 & 0.37 & 0.12 & 3.03 & 0.00 \\
  vegcov5 & 0.14 & 0.13 & 1.06 & 0.29 & 0.14 & 0.13 & 1.10 & 0.27 \\
  vegcov6$^*$  & -0.29 & 0.16 & -1.85 & 0.06 & -0.29 & 0.15 & -1.88 & 0.06 \\ %\hline
  \multicolumn{9}{c}{Detection $p$}\\ %\hline
  Intercept:1$^*$ & 1.07 & 0.30 & 3.60 & 0.00 & 1.09 & 0.30 & 3.67 & 0.00 \\
 Intercept:2$^*$  & 1.48 & 0.12 & 12.00 & 0.00 & 1.48 & 0.12 & 12.04 & 0.00 \\
  Intercept:3$^*$  & 2.27 & 0.30 & 7.70 & 0.00 & 2.26 & 0.30 & 7.62 & 0.00 \\
  vegcov1$^*$  & 0.56 & 0.09 & 6.06 & 0.00 & 0.55 & 0.09 & 6.02 & 0.00 \\
  vegcov2$^*$  & -0.25 & 0.09 & -2.79 & 0.01 & -0.25 & 0.09 & -2.76 & 0.01 \\
  vegcov3 & -0.10 & 0.09 & -1.16 & 0.24 & -0.09 & 0.09 & -1.05 & 0.30 \\
  vegcov4$^*$  & 0.18 & 0.08 & 2.37 & 0.02 & 0.18 & 0.07 & 2.34 & 0.02 \\
  vegcov5$^*$  & 0.10 & 0.08 & 1.34 & 0.18 & 0.11 & 0.07 & 1.46 & 0.14 \\
  vegcov6$^*$  & -0.10 & 0.09 & -1.18 & 0.24 & -0.11 & 0.09 & -1.27 & 0.20 \\
  time & -0.07 & 0.07 & -0.93 & 0.35 & -0.07 & 0.07 & -0.98 & 0.33 \\
  temp$^*$  & -0.24 & 0.08 & -3.09 & 0.00 & -0.24 & 0.08 & -3.11 & 0.00 \\
  cloud$^*$  & -0.13 & 0.07 & -1.90 & 0.06 & -0.13 & 0.07 & -1.93 & 0.05 \\
  julian$^*$  & -0.74 & 0.23 & -3.23 & 0.00 & -0.73 & 0.23 & -3.18 & 0.00 \\
   \hline
\end{tabular}
\end{tiny}
\label{tab-modelt}
\end{table}

\subsection{Data II}
A smaller data set is given on the website
James Peterson \footnote{\url{http://people.oregonstate.edu/~peterjam/occupancy_workshop/hands_on.html}}
that presents data on detections of brook trout collected via electrofishing in three 50 m sections of streams at 57 sites in the Upper Chattachochee
371 River basin, USA.
These data contained a site covariate, Elevation (Ele) and a time dependent covariate stream mean cross-sectional area (CSA).
These variables are on quite different scales. The average elevation was approximately 2861 and the mean cross-sectional area was less than 2.
We considered four models, just the site covariates, site covariates and time varying intercepts, site and survey covariates and
site and survey covariates with time varying intercepts. Using the default settings in {\tt occu} the estimates did not converge.
This was rectified by using the ``Nelder-Mead'' method set to a maximum of 2000 iterations. The two-stage estimator had no such problems. The estimates
for the unstandardised data are in Table \ref{tab-trout-us} (a). The estimates are generally similar. For the standardised data, {\tt occu} with the default options did converge.
The results are given in Table \ref{tab-trout-us} (b). The estimates are again quite similar.

\begin{table}[ht]
    \caption{Occupancy and detection estimates for full likelihood and two-stage approaches for the (a) unstandardised and (b) standardised brook trout data. For each covariate, we report its: estimate (Estimate), standard error (se), Student's \emph{t}-statistic ($t$), and $p$-value ($p$). Occupancy for the two-stage approach estimated with IWLS method.} 
%    , or direct maximisation if IWLS did not converge after 200 iterations.}}
\centering
\begin{tiny}
\begin{tabular}{rrrrrrrrr}
    \hline
      &\multicolumn{4}{c}{Full Likelihood}&\multicolumn{4}{c}{Two-stage}\\
     %\hline
    Parameter&  Estimate & se & $t$&$p$ &  Estimate & se & $t$&$p$ \\
    \hline
    \multicolumn{9}{l}{(a) Unstandardised}\\
     \multicolumn{9}{c}{Occupancy $\psi$}\\ %\hline
%  \hline
Intercept & -3.9716 & 0.6858 & -5.7914 & 0.0000 & -4.0452 & 1.1218 & -3.6060 & 0.0003 \\
  Ele  & 0.0013 & 0.0003 & 4.5338 & 0.0000 & 0.0013 & 0.0004 & 3.6441 & 0.0003\\ %\hline
   \multicolumn{9}{c}{Detection $p$}\\ %\hline
 Intercept & 0.0580 & 0.7352 & 0.0788 & 0.9372 & -0.1609 & 1.2397 & -0.1298 & 0.8968 \\
  Ele & 0.0004 & 0.0002 & 1.9697 & 0.0489 & 0.0004 & 0.0003 & 1.2516 & 0.2107 \\
  CSA & -0.8325 & 0.2822 & -2.9503 & 0.0032 & -0.7438 & 0.2873 & -2.5888 & 0.0096 \\
   \hline %\hline
    \multicolumn{9}{l}{(b) Standardised}\\
    \multicolumn{9}{c}{Occupancy $\psi$}\\ %\hline
  Intercept & -0.19 & 0.36 & -0.52 & 0.60 & -0.34 & 0.32 & -1.04 & 0.30 \\
    Ele & 1.53 & 0.45 & 3.42 & 0.00 & 1.48 & 0.40 & 3.71 & 0.00 \\
      %\hline

       \multicolumn{9}{c}{Detection $p$}\\
    %\hline
    Intercept & -0.14 & 0.35 & -0.38 & 0.70 & -0.16 & 0.36 & -0.44 & 0.66 \\
    Ele & 0.36 & 0.35 & 1.04 & 0.30 & 0.43 & 0.37 & 1.18 & 0.24 \\
    CSA & -0.82 & 0.28 & -2.97 & 0.00 & -0.80 & 0.28 & -2.81 & 0.00 \\ \hline
\end{tabular}
\end{tiny}
\label{tab-trout-us}
\end{table}

\clearpage

\section{Discussion}\label{sec-disc}

In \citet{kara17} we examined the two-stage approach for the homogeneous occupancy model.
Here we examined the two-stage approach for the heterogeneous occupancy model where the occupancy and detection probabilities now depend on covariates that may vary between sites and over time. 

In our applications here the two-stage estimator gave similar estimates to the full maximum likelihood with the {\tt occu} function in package {\tt unmarked}.
For the large standardised data set first considered the estimates from the two methods were very similar.
The two-stage method has advantages in model selection as the dimension of the space to be searched can be enormously reduced.
By considering two smaller dimensional parameter spaces and using IWLS in both stages it is also numerically more stable so that standardisation
is less important. It also gives access to the {\tt VGAM} methodology in estimating detection. We defer further exploration elsewhere. 

The default in the {\tt occu} function in the {\tt unmarked} package uses a vector of zeroes as starting values and if the algorithm does not converge
suggests the user provides starting values. However, there is no guidance on how to find suitable values. A simulation study showed that {\tt occu} may give extreme estimates of occupancy parameters when the two-stage does not.  

If there are too few redetections the two-stage estimator can fail. This results from conditioning on at least one detection. However, when the main focus is on estimating occupancy, the occupancy probability appears to be relatively insensitive for small changes in detection probability \citep{kara17}.

\clearpage

\section*{References}

\bibliography{July-2018}
\bibliographystyle{apalike}

%\clearpage

\appendix

    \section{Conditional Likelihood using {\tt VGAM}}\label{vgam}

    The R package {\tt VGAM} \citep{yee10} is a powerful and flexible package that fits models to vector responses.
    As such, at first glance it can be overwhelming. However, its handling of time dependent covariates makes it preferable to
    writing one's own functions.
    Here we give a description of how it can be used to fit some common models
    to detections using conditional likelihood in the first stage of our approach.

    \subsection{Fitting Time Independent Covariates for Detection}\label{vgam-tid}

    In our two-stage approach the conditional likelihood fits the model for detection to data from the sites where there was at least one detection. This can be done the {\tt posbinomial} family in the {\tt VGAM} function {\tt vglm}. Firstly the data is reduced to those sites where there was at least one detection.
    When there is no time dependence, computing the estimates using {\tt vglm} is straightforward.
    We illustrate this for the data of \citet{hutchinson15,dryad_t40f2}
         as these data contain both site and visit (i.e. time dependent) covariates.
    For these data, the data frame {\tt data} is a reduced data frame that contains data from the sites where occupancy was detected. The variable $Y$ is the number of times the species was detected at each occupied site, $\tau$ is the number of visits to each site, and site covariates are vegcov1,vegcov2,\dots,vegcov6. See Figure \ref{fig-1} for selected output ($\tau = 3$ in this example). The parameter estimates may then be used in the second stage of the analysis.

\begin{figure}
\begin{knitrout}\small
\definecolor{shadecolor}{rgb}{0.969, 0.969, 0.969}\color{fgcolor}\begin{kframe}
\begin{verbatim}
> V.out=vglm(cbind(Y,3-Y)~vegcov1+vegcov2+vegcov3+vegcov4
        +vegcov5+vegcov6,
    family=posbinomial(omit.constant=TRUE),data=data)
> coef(V.out)
# (Intercept)     vegcov1     vegcov2     vegcov3     vegcov4
#   1.5590909   0.5493825  -0.2512287  -0.1048756   0.1656597
#  vegcov5     vegcov6
#  0.1186192   -0.1277806
\end{verbatim}
\end{kframe}
\end{knitrout}
\caption{Fitting the detection model for time homogeneous covariates.}
\label{fig-1}
\end{figure}

     With the  univariate response $Y$, the implementation is very similar to {\tt glm}.
     The term {\tt omit.constant=TRUE} does not affect the fitting but removes the constant terms from the computation of the AIC.
These estimates may then be input into the second stage procedure to estimate parameters associated with the occupancy model.

    \subsection{Fitting Time Dependent Covariates for Detection}\label{vgam-t}

    Time dependent models for detection may  be fitted to data using the {\tt posbernoulli.t} family in the
    {\tt vglm} function in {\tt VGAM}.
    Fitting these models is more complex as many more models are available and the response consists of the detections on each visit to the
    site and is hence  multivariate.
    We again use the data from  \citet{hutchinson15,dryad_t40f2}. The time dependent covariates are time, temp, cloud and julian measured for each visit to each site.
    These are included in the data data frame as time1, time2, \dots julian2, julian3.
    With a vector valued response there is the possibility that the coefficient associated with a covariate may change with the visit, so the associated modelling
    and hence the functions to fit the models are more complex. See  \citet[\S6.3]{yee10} for a worked example in a capture-recapture context.
     In Figure \ref{fig-2} we first fit a simple model with time dependent intercepts and the
    relationship with the site covariates remains independent of time. This was specified through the {\tt parallel.t} argument to the {\tt posbernoulli.t}
    family. Note that {\tt parallel.t=FALSE$\sim$1} is the default for the {\tt  posbernoulli.t} family but  for clarity we explicitly incorporate it in Figure \ref{fig-2}.

\begin{figure}
\begin{knitrout}\small
\definecolor{shadecolor}{rgb}{0.969, 0.969, 0.969}\color{fgcolor}\begin{kframe}
\begin{verbatim}
> V.out=vglm(cbind(survey1,survey2,survey3)
       ~ vegcov1+vegcov2+vegcov3+vegcov4+vegcov5+vegcov6,
     family=posbernoulli.t(parallel.t=FALSE~1), data=data)
> coef(V.out)
#  (Intercept):1 (Intercept):2 (Intercept):3  vegcov1    vegcov2
#  1.8583766     1.5130892     1.3527893      0.5515551  -0.2522520
#   vegcov3    vegcov4    vegcov5    vegcov6
#  -0.1052631  0.1663444  0.1190595  -0.1282785
\end{verbatim}
\end{kframe}
\end{knitrout}
\caption{Fitting a model with time dependent intercepts.}
\label{fig-2}
\end{figure}

Incorporating time dependent covariates is a little more complex and requires use of the {\tt xij} and {\tt form2} arguments in {\tt VGAM}.
The {\tt form2} argument is straightforward. It gives all the variables in the model and needs to be included if {\tt xij} is used. The  {\tt xij} argument specifies that covariates have different values
at different visits. To implement it is necessary to construct a new variable, for example {\tt time.tij}, for each time dependent variable in the model and incorporate them in
the data frame. In our case this gives four new variables, {\tt time.tij}, {\tt temp.tij}, {\tt cloud.tij} and {\tt julian.tij} in Figure \ref{fig-3}.

    \begin{figure}
\begin{knitrout}\small
\definecolor{shadecolor}{rgb}{0.969, 0.969, 0.969}\color{fgcolor}\begin{kframe}
\begin{verbatim}
> V.out=vglm(cbind(survey1,survey2,survey3)
  ~vegcov1+vegcov2+vegcov3+vegcov4
    +vegcov5+vegcov6+time.tij+temp.tij+cloud.tij+julian.tij,
 data=Data.all,
 xij=list(time.tij~time1+time2+time3-1,temp.tij~temp1+temp2+temp3-1,
 cloud.tij~cloud1+cloud2+cloud3-1,julian.tij~julian1+julian2
    +julian3-1),
 family=posbernoulli.t(parallel.t=FALSE~0),
 form2=~vegcov1+vegcov2+vegcov3+vegcov4+vegcov5+vegcov6+time.tij
    +temp.tij+cloud.tij+julian.tij+time1+time2+time3+temp1+temp2
    +temp3+cloud1+cloud2+cloud3+julian1+julian2+julian3)
> coef(V.out)
# (Intercept)  vegcov1    vegcov2     vegcov3      vegcov4
#  1.60651791  0.54525171 -0.24061702 -0.08727207  0.16955603
# vegcov5  vegcov6   time.tij  temp.tij  cloud.tij julian.tij
# 0.108527 -0.112085 -0.069456 -0.238605 -0.161028 -0.264658
\end{verbatim}
\end{kframe}
\end{knitrout}
    \caption{Fitting a model using with time varying covariates but constant intercept for the two-stage approach in {\tt vglm}.}
    \label{fig-3}
    \end{figure}

    \section{Iterative Weighted Least Squares}\label{sec-iwls}

    The potential instability of the maximum likelihood estimates when computed using numerical optimization, through the function {\tt optim} in R
    motivated us to develop an iterative weighted least squares (IWLS) approach. This is quite straightforward for the logistic model in our two-stage approach.

Recall $E(w_s)=\eta_s=\theta_s\psi_s$. Let $\boeta=(\eta_1,\dots,\eta_S)^T$, $\bw=(w_1,\dots,w_S)^T$
and the $ q_0 \times S$ matrix $X$ has $s$th column $x_s.$
Then, as $\theta_s$ is not a function of $\balpha$, maximising the partial log-likelihood (\ref{eq-log-PL}) is equivalent to maximising
$
    \ell(\boeta)=\sum_{s=1}^S \left\{(1-w_s)\log(1-\eta_s)+w_s \log(\eta_s) \right\}
$.
Then, with $V={\rm diag}\{(1-\boeta)\boeta\}$ we have
$
    \bu(\boeta)={\partial \ell(\boeta)}/{\partial \boeta}=V^{-1}(\bw-\boeta)
$. Let $\gamma_s = \bx_s^T\balpha$ and $\bgamma=(\gamma_1,\dots,\gamma_S)=X\balpha$.
Now,
$
    {\partial \eta_s}/{\partial \gamma_s}=\theta_s\psi_s(1-\psi_s)$
and
$
    {\partial  \gamma_s}/{\partial \balpha}=\bx_s
$
so that
$
    {\partial  \eta_s}/{\partial \balpha}=\theta_s\psi_s(1-\psi_s)\bx_s$.
That is,
$
    {\partial \boeta^T}/{\partial \balpha}= X U
$
where
$
    U={\rm diag}\{\theta_s\psi_s(1-\psi_s)\}
$
or the partial score equations may be written as
$
    \bu(\balpha)=\left({\partial \boeta^T}/{\partial \balpha}\right)\bu(\boeta)=X U V^{-1}(\bw-\boeta(\balpha))$.
The expected conditional Fisher information is then
$
    J(\balpha)=-E\left({\partial \bu(\balpha)}/{\partial \balpha^T} \right)=X U V^{-1}U X^T
$.
Recall $\bZ=UX\balpha^{(k)}+ \bw-\boeta(\balpha^{(k)})$. Then
\begin{align*}
\balpha^{(k+1)}&\approx\balpha^{(k)}+J(\balpha)^{-1}\bu(\balpha^{(k)})\\
&=\balpha^{(k)}+\left(X U V^{-1}U X^T \right)^{-1}X U V^{-1}(\bw-\boeta(\balpha^{(k)}))\\
&=\left(X U V^{-1}U X^T \right)^{-1}X U V^{-1} \left(UX\balpha^{(k)}+ \bw-\boeta(\balpha^{(k)})\right)\\
&=\left(X U V^{-1}U X^T \right)^{-1}X U V^{-1}  \bZ.
\end{align*}
This gives the iterative procedure of Section \ref{sec-iter1}.

\section{Derivation of the Standard Errors}\label{sec-glmp}

We outline the proof  of
\eqref{eq-se} for the linear model with logistic link. Let $\balpha_0$ and $\bbeta_0$ denote the true values of $\balpha$ and
$\bbeta$ and let $\widehat\bbeta$ be a consistent
estimator of $\bbeta$. Here this will be the conditional likelihood estimator of $\bbeta$
but our results are more general than that.
We suppose that for a $q_o \times q_p$ matrix $B(\balpha,\bbeta)$,
and a $q_o \times q_o$ matrix $A(\balpha,\bbeta)$,
\begin{align*}
S^{-1} \widetilde B(\balpha_0,\bbeta_0)=S^{-1}\frac{\partial Q(\balpha_0,\bbeta_0)}{\partial
\bbeta_0^T}
&\rightarrow B(\alpha_0,\beta_0),\\
-S^{-1}\frac{\partial Q(\alpha_0,\beta_0)}{\partial
\alpha_0^T} &\rightarrow A(\alpha_0,\beta_0)
\end{align*}
and that the central limit theorem is applicable so that
\begin{equation}
S^{-1/2} Q(\alpha_0,\beta_0) \buildrel d \over \longrightarrow N(0,\Sigma_Q).
\end{equation}
We also suppose that the estimators $\widehat\bbeta$ arising from the
first stage are consistent and satisfy
$
S^{1/2}(\widehat\bbeta-\bbeta_0) \buildrel d \over \longrightarrow N(0,\Sigma_\beta)
$ for some $q \times q$ matrix $\Sigma_\beta$. That is, ${\rm Var}(\whbeta)=\Sigma_\beta/S$.
Note that using {\tt vglm} to estimate $\bbeta$ using conditional likelihood yields an estimate
of ${\rm Var}(\whbeta\vert O)$. Then we approximate $\Sigma_{\bbeta}$ by $S {\rm Var}(\whbeta\vert O)$.
Finally, we suppose that the partial score functions for $\balpha$ are
uncorrelated with those for $\bbeta$. We have noted in \S 
\ref{sec-two-stage} that this holds for the likelihood conditional on at least one detection at a site.

The log-partial likelihood is \\
$
\ell(\balpha,\bbeta)=\sum_{s=1}^S\left\{ (1-w_s)\log(1-\psi_s(\balpha)\theta_s)+w_s \log(\psi_s(\balpha))\right\},
$
so that \\
$
Q(\balpha,\bbeta)={\partial \ell(\balpha,\bbeta)}/{\partial
\alpha}=\sum_{s=1}^S x_s
{\{w_s-\psi_s(\alpha)\theta_s\}(1-\psi_s(\alpha))}/\{1-\psi_s(\alpha)\theta_s\}.
$
The first order expansion of $Q(\widehat\balpha,\widehat\bbeta)$ about $\balpha_0$ yields
\[
S^{1/2}\left(\widehat\balpha(\widehat\bbeta)-\balpha_0\right)= \left\{ S^{-1} I(\balpha_0,\widehat\bbeta)\right\}^{-1}S^{-1/2} Q(\balpha_0,\widehat\bbeta),
\]
and that of $ Q(\balpha_0;\widehat\bbeta)$ about $\bbeta_0$ yields
\[
S^{-1/2} Q(\balpha_0;\widehat\bbeta)\approx S^{-1/2}
Q(\balpha_0,\bbeta_0)
+S^{-1}\widetilde B(\balpha_0,\bbeta_0)S^{1/2}(\widehat\bbeta-\bbeta_0)
\]
which together give
\[
S^{1/2}(\widehat\balpha(\widehat\bbeta)-\balpha_0) \approx
A(\balpha_0,\bbeta)^{-1}\left\{ S^{-1/2} Q(\balpha_0,\bbeta_0)
+ B(\balpha,\bbeta_0)S^{1/2}(\widehat\bbeta-\bbeta_0)
\right\}.
\]
The central limit theorem and recalling that the
partial score functions are uncorrelated then gives
\[
S^{1/2}(\whalpha(\widehat\bbeta)-\alpha_0) \sim
N_p\left(0,A(\balpha_0,\bbeta_0)^{-1}\left\{\Sigma_Q+B(\balpha_0,\bbeta_0)\Sigma_\bbeta B(\balpha_0,\bbeta_0)^T\right\}A(\balpha_0,\bbeta_0)^{-T}\right),
\]
where
\[
\Sigma_Q=S^{-1}{\rm Var}\left(Q(\balpha_0,\bbeta_0) \right)=S^{-1}E\left(I(\balpha_0,\bbeta_0) \right).
\]
That is
$
{\rm Var}\{\whalpha(\whbetas)\}=S^{-1} \left\{A(\alpha_0,\beta_0)^{-1} + A(\alpha_0,\beta_0)^{-1} B(\alpha_0,\beta_0)\Sigma_\beta B(\alpha_0,\beta_0)^TA(\alpha_0,\beta_0)^{-T}\right\}$.
To estimate the standard errors, recall
$
I(\alpha,
\beta)  = -{\partial Q(\alpha,\beta)}/{\partial \alpha^T}$ yielding \eqref{eq-I}.

Next we determine $\widetilde B(\alpha,\beta)$ in the time homogeneous case. As
\begin{equation}
\frac{\partial Q(\alpha,\beta)}{\partial
\theta_s}=-\frac{x_s\psi_s(1-\psi_s)(1-w_s)}{(1-\psi_s\theta_s)^2},\label{eq-B}
\end{equation}
and $\theta_s=1-(1-p_s)^\tau$,
the chain rule gives,
\[
q_s=\frac{\partial Q(\alpha,\beta)}{\partial p_s}= -\frac{x_s\psi_s(1-\psi_s)(1-w_s) \tau(1-p_s)^{\tau-1}}{(1-\psi_s\theta_s)^2}.
\]
As $\partial p_s/\partial \beta=p_s(1-p_s)u_s$ we then see that
\begin{equation*}
\frac{\partial Q(\alpha,\beta)}{\partial \beta^T}=\widetilde B(\alpha,\beta)=-\sum_{s=1}^S
\frac{x_su_s^T\psi_s(1-\psi_s)(1-w_s)
\tau(1-p_s)^{\tau}p_s}{(1-\psi_s\theta_s)^2}.%\label{eq-B}
\end{equation*}
As $\Sigma_\beta=S{\rm Var}(\whbetas)=S V_\beta$ where
$V_\beta$ is the covariance matrix of $\whbeta$ and it is easily seen that
${\rm Var}\{\whalpha(\whbetas)\}$
reduces to \eqref{eq-se}.

In the time heterogeneous case $\theta_s=1-\prod_{j=1}^\tau(1-p_{sj})$.
Hence
	for the logistic model,
	\[
	\frac{\partial \theta_s}{\partial \beta}= \sum_{j=1}^\tau \prod_{k
	  \ne j} (1-p_{sk})\frac{\partial p_{sj}}{\partial \beta}
	= \sum_{j=1}^\tau \prod_{k=1}^\tau (1-p_{sk}) p_{sj}u_{sj}
	=(1-\theta_s) \sum_{j=1}^\tau p_{sj} u_{sj}.
	\]
	As (\ref{eq-B}) still holds  this yields
	the modification of the expression  \eqref{eq-se} for the standard errors
	as given in \S \ref{sec-td0}.

    \section{Fitting the Full Likelihood with {\tt occu}}\label{sec-occu}

    The use of {\tt occu} is well documented. Here, we briefly describe its use as it handles time varying (or time dependent) covariates differently to {\tt vglm}.
    To fit our full model using {\tt occu} we first construct a matrix of factors, {\tt Visit},
    corresponding to the three visits. We then construct a list {\tt Obs} that contains data frames of
    the time varying covariates. This is then converted into an {\tt unmarkedFrameOccu} object, {\tt D}.
    The model is then fitted to the data, as shown in Figure \ref{fig-4}. Thus in either the {\tt vglm} or {\tt occu} approaches there is
    an initial data manipulation step requiring construction of an appropriately structured data frame, then the fitting to data.
    With {\tt vglm} there is then a second step to estimate the occupancy model.

          \begin{figure}
      \begin{knitrout}\small
      \definecolor{shadecolor}{rgb}{0.969, 0.969, 0.969}\color{fgcolor}\begin{kframe}
\begin{verbatim}
> Visit=matrix(as.factor(c(rep("a",656),rep("b",656),rep("c",656))),
    ncol=3)
> Obs=list(time=as.data.frame(Model.out@T.ij[,c(1,5,9)]),
     temp=as.data.frame(Model.out@T.ij[,c(2,6,10)]),
     cloud=as.data.frame(Model.out@T.ij[,c(3,7,11)]),
     julian=as.data.frame(Model.out@T.ij[,c(4,8,12)]),
     Visit=as.data.frame(Visit))
> D=unmarkedFrameOccu(y=Model.out@Detect,
     siteCovs=as.data.frame(Model.out@X[,-1]),obsCovs=Obs)
> O.5.out=occu(~Visit+vegcov1+vegcov2+vegcov3+vegcov4+vegcov5+vegcov6
     +time+temp+cloud+julian-1~vegcov1+vegcov2+vegcov3+vegcov4+vegcov5
     +vegcov6,data=D,engine=c("C"))
> O.5.out@estimates
      \end{verbatim}
      \end{kframe}
      \end{knitrout}
          \caption{Fitting a model with {\tt occu} for time varying covariates on the full model.}
          \label{fig-4}
          \end{figure}

\end{document}